\documentstyle[manuscript,aps,prl]{revtex}
\input{epsf}
\begin{document}

\title{Entangled-photon ellipsometry}

\author{Ayman F. Abouraddy, Kimani C. Toussaint, Jr., Alexander V. Sergienko, Bahaa E. A. Saleh, and Malvin C. Teich}
\address{Quantum Imaging Laboratory, Departments of Electrical $\&$ Computer Engineering and Physics, Boston University, Boston, MA $02215-2421$}
\maketitle

\begin{abstract}Performing reliable measurements in optical metrology, such as
those needed in ellipsometry, requires a calibrated source and
detector, or a well-characterized reference sample. We present a
novel interferometric technique to perform reliable ellipsometric
measurements. This technique relies on the use of a non-classical
optical source, namely polarization-entangled twin photons
generated by spontaneous parametric downconversion from a
nonlinear crystal, in conjunction with a coincidence-detection
scheme. Ellipsometric measurements acquired with this scheme are
absolute, i.e., they require neither source nor detector
calibration, nor do they require a reference.
\end{abstract}

\section{INTRODUCTION}

A question that arises frequently in metrology is the following:
how does one measure reliably the reflection or transmission
coefficient of an unknown sample? The outcome of such a
measurement depends on the reliability of both the source and the
detector used to carry out the measurements. If they are each
absolutely calibrated, such measurements would be trivial. Since
such ideal conditions are never met in practice, and since high
precision measurements are often required, a myriad of
experimental techniques, such as null and interferometric
approaches, have been developed to circumvent the imperfections of
the devices involved in these measurements.

One optical metrology setting in which high-precision measurements
are a necessity is ellipsometry \cite{PD,RA,HT,AR,AW,MM}, in which
the polarization of light is used to study thin films on
substrates, a technique established more than a hundred years ago
\cite{PD,AR,AW}. Ellipsometers have proven to be an important
metrological tool in many arenas ranging from the semiconductor
industry to biomedical applications. To carry out \textit{ideal}
ellipsometry one needs a perfectly calibrated source and detector.
Various approaches, such as null and interferometric techniques
have been commonly used in ellipsometers \cite{RA,MM} to approach
this ideal. Section 2 of this paper describes the basic
requirements for ideal ellipsometry and reviews some of the more
common techniques that have been used in conjunction with
available detectors and sources.

In this paper we propose a novel technique for obtaining reliable
ellipsometric measurements based on the use of twin photons
produced by the process of spontaneous optical parametric
downconversion (SPDC)\cite{KL67,Harris67,Gia68,Kle68,DK1}. This
source has been used effectively in studies of the foundations of
quantum mechanics \cite{AZ99,EF} and in applications in quantum
metrology \cite{Kly77,DB2,AA2,AM} ; quantum information
processing, such as quantum cryptography \cite{Ekert92,AS,Jen00};
quantum teleportation \cite{DB1,Bos98}; and quantum imaging
\cite{BS2,Nasr01,Abou01}. We extend the use of this non-classical
light source to the field of ellipsometry.

In Section 3, we propose two different experimental
implementations of twin-photon ellipsometry. The first makes use
of a twin-photon interferometer that has been previously used for
testing the foundations of quantum mechanics. The second technique
makes direct use of polarization-entangled photon pairs emitted
via SPDC. This approach effectively comprises an interferometric
ellipsometer, although none of the optical elements usually
associated with constructing an interferometer are utilized.
Instead, polarization entanglement itself is harnessed to perform
interferometry, and to achieve ideal ellipsometry. The inherent
limitations of the first technique are eliminated in the second.

\section{IDEAL ELLIPSOMETRY}

In an ideal ellipsometer, the light emitted from a reliable
optical source is directed into an unknown optical system (which
may simply be an unknown sample that reflects the impinging light)
and thence into a reliable detector. The practitioner keeps track
of the emitted and detected radiation, and from this bookkeeping
(s)he can infer information about the optical system. This device
may be used as an ellipsometer if the source can emit light in any
specified state of polarization. The sample is characterized by
two parameters: $\psi$ and $\Delta$. The quantity $\psi$ is
related to the magnitude of the ratio of the sample's
eigenpolarization complex reflection coefficients, $\tilde{r}_{1}$
and $\tilde{r}_{2}$, via $\tan\psi=|\tilde{r}_{1}/\tilde{r}_{2}|$;
$\Delta$ is the phase shift between them \cite{RA}.

Because of the high accuracy required in measuring these
parameters, an ideal ellipsometric measurement would require
absolute calibration of both the source and the detector. Since
this is not attainable in practical settings, ellipsometry makes
use of a myriad of experimental techniques developed to circumvent
the imperfections of the involved devices. The most common
techniques are null and interferometric ellipsometry.

In the traditional null ellipsometer \cite{RA}, depicted in
Fig.~\ref{null}, the sample is illuminated with a beam of light
that can be prepared in any state of polarization. The reflected
light, which is generally elliptically polarized, is then
analyzed. The polarization of the incident beam is adjusted to
compensate for the change in the relative amplitude and phase,
introduced by the sample, between the two eigenpolarizations, so
that the resulting reflected beam is linearly polarized. If passed
through an orthogonal linear polarizer, this linearly polarized
beam will yield a null (zero) measurement at the optical detector.
The null ellipsometer does not require a calibrated detector since
it does not measure intensity, but instead records a null. The
principal drawback of null measurement techniques is the need for
a reference to calibrate the null, for example to find its initial
location (the rotational axis of reference at which an initial
null is obtained) and then to compare this with the subsequent
location upon inserting the sample into the apparatus. Such a
technique thus alleviates the problem of an unreliable source and
detector, but necessitates the use of a reference sample. The
accuracy and reliability of all measurements depend on our
knowledge of this reference sample. In this case, the measurements
are a function of $\psi$, $\Delta$, and the parameters of the
reference sample.

Another possibility is to perform ellipsometry that employs an
interferometric configuration in which the light from the source
follows more than one path, usually created via beam splitters,
before reaching the detector. The sample is placed in one of those
paths. We can then estimate the efficiency of the detector
(assuming a reliable source) by performing measurements when the
sample is removed from the interferometer. This configuration thus
alleviates the problem of an unreliable detector, but depends on
the reliability of the source and suffers from the drawback of
requiring several optical components (beam splitters, mirrors,
etc.). The ellipsometric measurements are a function of $\psi$,
$\Delta$, source intensity, and the parameters of the optical
elements. The accuracy of the measurements are therefore limited
by our knowledge of the parameters characterizing these optical
components. The stability of the optical arrangement is also of
importance to the performance of such a device.

\section{TWIN-PHOTON ELLIPSOMETRY}

All classical optical sources (including ideal
amplitude-stabilized lasers) suffer from unavoidable quantum
fluctuations even if all other extraneous noise sources are
removed. Fluctuations in the photon number can only be eliminated
by constructing a source that emits non-overlapping wave packets,
each of which contains a fixed photon number. Such sources have
been investigated, and indeed sub-Poisson light sources have been
demonstrated \cite{Teich85,Teich88,Teich90}.

One such source may be readily realized via the process of
spontaneous parametric downconversion (SPDC) from a second-order
nonlinear crystal (NLC) when illuminated with a monochromatic
laser beam (pump) \cite{DK1}. A portion of the pump photons
disintegrate into photon pairs. The two photons that comprise the
pair, known as signal and idler, are highly correlated since they
conserve the energy (frequency-matching) and momentum
(phase-matching) of the parent pump photon.

In type-II SPDC the signal and idler photons have orthogonal
polarizations, one extraordinary and the other ordinary. These two
photons emerge from the NLC with a relative time delay due to the
birefringence of the NLC \cite{PK}. Passing the pair through an
appropriate birefringent material of suitable length compensates
for this time delay. This temporal compensation is required for
extracting $\psi$ and $\Delta$ from the measurements; we show
subsequently that when compensation is not employed one may obtain
$\psi$ but not $\Delta$.

The signal and idler may be emitted in two different directions, a
case known as non-collinear SPDC, or in the same direction, a case
known as collinear SPDC. In the former situation, the SPDC state
is polarization entangled; its quantum state is described by
\cite{PK}
\begin{equation}\label{entangled-state}
|\Psi\rangle=\frac{1}{\sqrt{2}}(|HV\rangle+|VH\rangle),
\end{equation}
where {\it H} and {\it V} represent horizontal and vertical
polarizations, respectively \cite{AS2}. It is understood that the
first polarization indicated in a ket is that of the signal photon
and the second is that of the idler. Such a state may not be
written as the product of states of the signal and idler photons.
Although Eq. (\ref{entangled-state}) represents a pure quantum
state, the signal and idler photons considered separately are each
unpolarized \cite{UF,AA1}. The state represented in Eq.
(\ref{entangled-state}) assumes that there is no relative phase
between the two kets. Although the relative phase may not be zero,
it can, in general, be arbitrarily chosen by making small
adjustments to the NLC.

In the collinear case the SPDC state is in a polarization-product
state
\begin{equation}\label{separable}
|\Psi\rangle=|HV\rangle.
\end{equation}
Because this state is factorizable (i.e., it may be written as the
product of states of the signal and idler photons), it is not
entangled.

We first discuss a configuration based on the use of collinear
type-II SPDC, which we call a \textit{unentangled twin-photon
ellipsometer}. This configuration is introduced for pedagogical
reasons as a precursor to the configuration of principal interest
to us, called the \textit{entangled twin-photon ellipsometer},
which makes use of polarization-entangled photon pairs
(non-collinear type-II SPDC). Both arrangements are described
using a generalization of the Jones-matrix formalism appropriate
for twin-photon polarized beams.

\subsection{UNENTANGLED TWIN-PHOTON ELLIPSOMETER}

We now examine the use of collinear type-II SPDC in a standard
twin-photon polarization interferometer, previously used in
numerous experiments \cite{AS2}, and shown in
Fig.~\ref{polariz-int}. The twin photons, with the state shown in
Eq.\ (\ref{separable}), impinge on the input port of a
non-polarizing beam splitter, so that on 50$\%$ of the trials the
two photons are separated into the two output ports of the beam
splitter \cite{RC}. On the remainder of the trials, the two
photons emerge together from the beam splitter out of one of the
ports, but such cases do not contribute to coincidence
measurements and thus may be ignored. Photons emerging from the
one of the output ports of the beam splitter are directed to the
sample under test and are then directed to polarization analyzer
A$_{1}$ followed by single-photon detector D$_{1}$. Photons
emerging from the other output port are directed to polarization
analyzer A$_{2}$ followed by single-photon detector D$_{2}$. A
coincidence circuit registers the coincidence rate $N_{c}$ of the
detectors D$_{1}$ and D$_{2}$, which is proportional to the
fourth-order coherence function of the fields at the detectors
\cite{BS,RG}. In this section, we demonstrate how this unentangled
twin-photon polarization interferometer yields ellipsometric
measurements.

We first introduce a matrix formalism that facilitates the
derivation of the fields at the detectors. We begin by defining a
\textit{twin-photon Jones vector} that represents the field
operators of the signal and idler in two spatially distinct modes.
If $\hat{a}_{s}(\omega)$ and $\hat{a}_{i}(\omega')$ are the boson
annihilation operators for the signal-frequency mode $\omega$ and
idler-frequency mode $\omega'$, respectively, then the twin-photon
Jones vector of the field following the beam splitter is
\begin{equation}\label{jones-vect-beamsplit}
\hat{\textbf{J}}_{1} = \left( \matrix{
  j\,\{-\hat{\textbf{A}}_{s}(\omega)+\hat{\textbf{A}}_{i}(\omega')\}\cr
  \hat{\textbf{A}}_{s}(\omega)+\hat{\textbf{A}}_{i}(\omega')\cr} \right),
\end{equation}
where $\hat{\textbf{A}}_{s}(\omega)=\hat{a}_{s}(\omega) \left(
\matrix{
  1\cr
  0\cr
} \right)$ and $\hat{\textbf{A}}_{i}(\omega')=\hat{a}_{i}(\omega')
\left(
  \matrix
  {0\cr
  1\cr} \right)$ \cite{LesHouches}. The vectors $\left(
  \matrix{
  1\cr
  0\cr
} \right)$ (horizontal) and $\left(
  \matrix{
  0\cr
  1\cr
} \right)$ (vertical) are the familiar Jones vectors representing
orthogonal polarization states \cite{SalehTeich}. The operators
$\hat{\textbf{A}}_{s}(\omega)$ and $\hat{\textbf{A}}_{i}(\omega')$
thus are annihilation operators that include the vectorial
polarization information of the field mode. The first element in
$\hat{\textbf{J}}_{1}$,
$j\,\{-\hat{\textbf{A}}_{s}(\omega)+\hat{\textbf{A}}_{i}(\omega')\}$
represents the annihilation operator of the field in beam 1, which
is a superposition of signal and idler field operators. The second
element in $\hat{\textbf{J}}_{1}$,
$\hat{\textbf{A}}_{s}(\omega)+\hat{\textbf{A}}_{i}(\omega')$, is
the annihilation operator of the field in beam 2.

We now define a\textit{ twin-photon Jones matrix} that represents
the action of linear deterministic optical elements, placed in the
two beams, on the polarization of the field as follows:
\begin{equation}\label{gjones-matrix}
\textbf{T}=
  \left(\matrix{
    \textbf{T}_{11} & \textbf{T}_{12}\cr
    \textbf{T}_{21} & \textbf{T}_{22}\cr
  }\right),
\end{equation}
where $\textbf{T}_{kl} \:(k,l=1,2)$ is the familiar $2\times2$
Jones matrix that represents the polarization transformation
performed by a linear deterministic optical element. The indices
refer to the spatial modes of the input and output beams. For
example, $\textbf{T}_{11}$ is the Jones matrix of an optical
element placed in beam 1 whose output is also in beam 1, whereas
$\textbf{T}_{21}$ is the Jones matrix of an optical element placed
in beam 1 whose output is in beam 2, and similarly for
$\textbf{T}_{12}$ and $\textbf{T}_{22}$. In most cases, when an
optical element is placed in beam 1 and another in beam 2,
$\textbf{T}_{12}=\textbf{T}_{21}=\textbf{0}$. An exception is,
e.g., a beam splitter with beams 1 and 2 incident on its two input
ports, or other optical components that mix the spatial modes of
the two beams. The twin-photon Jones matrix $\textbf{T}$
transforms a twin-photon Jones vector $\hat{\textbf{J}}_{1}$ into
$\hat{\textbf{J}}_{2}$ according to
$\hat{\textbf{J}}_{2}=\textbf{T}\hat{\textbf{J}}_{1}$.

Applying this formalism to the arrangement in
Fig.~\ref{polariz-int}, assuming that beams 1 and 2 impinge on the
two polarization analyzers A$_{1}$ and A$_{2}$ directly (in
absence of the sample), the twin-photon Jones matrix is given by
\begin{equation}\label{gjones-matrix-polarizer}
\textbf{T}_{p}=
  \left(\matrix{
    \textbf{P}(-\theta_{1}) & \textbf{0}\cr
    \textbf{0} & \textbf{P}(\theta_{2})\cr
  }\right),
\end{equation}
where $\textbf{P}(\theta)=
  \left(\matrix{
    \cos^{2}\theta & \cos\theta\sin\theta\cr
    \cos\theta\sin\theta & \sin^{2}\theta\cr
  }\right),$ and $\theta_{1}$ and $\theta_{2}$ are
the angles of the axes of the analyzers with respect to the
horizontal direction. In this case the twin-photon Jones vector
following the analyzers is therefore
\begin{eqnarray}\label{jones-vect-analyzers}
&\hat{\textbf{J}}_{2} &= \textbf{T}_{p}\hat{\textbf{J}}_{1}=\left(
\matrix{
  j\,\textbf{P}(-\theta_{1})\{-\hat{\textbf{A}}_{s}(\omega)+\hat{\textbf{A}}_{i}(\omega')\}\cr
  \textbf{P}(\theta_{2})\{\hat{\textbf{A}}_{s}(\omega)+\hat{\textbf{A}}_{i}(\omega')\}\cr
} \right)\cr\nonumber\\ && =\left( \matrix{
  j\{-\cos\theta_{1}\hat{a}_{s}(\omega)+\sin\theta_{1}\hat{a}_{i}(\omega')\}\left(
  \matrix{
  \cos\theta_{1}\cr
  -\sin\theta_{1}\cr
} \right)\cr
  \{\cos\theta_{2}\hat{a}_{s}(\omega)+\sin\theta_{2}\hat{a}_{i}(\omega')\}\left(
  \matrix{
  \cos\theta_{2}\cr
  \sin\theta_{2}\cr
} \right)
} \right).
\end{eqnarray}

Using the twin-photon Jones vector $\hat{\textbf{J}}_{2}$ one can
obtain expressions for the fields at the detectors. The
positive-frequency components of the field at detectors D$_{1}$
and D$_{2}$, denoted $\hat{\textbf{E}}_{1}^{+}$ and
$\hat{\textbf{E}}_{2}^{+}$ respectively, are given by
\begin{eqnarray}\label{field1-nosample}
&\hat{\textbf{E}}_{1}^{+}(t)&=j\{-\cos\theta_{1}\int d\omega\,
e^{-j\omega t}\hat{a}_{s}(\omega)+\;\sin\theta_{1}\int
d\omega'\,e^{-j\omega't}\hat{a}_{i}(\omega')\} \left(
  \matrix{
  \cos\theta_{1}\cr
  -\sin\theta_{1}\cr
} \right),
\end{eqnarray}

\begin{eqnarray}\label{field2-nosample}
&\hat{\textbf{E}}_{2}^{+}(t)&=\{\cos\theta_{2}\int d\omega\,
e^{-j\omega t}\hat{a}_{s}(\omega)+\;\sin\theta_{2}\int
d\omega'\,e^{-j\omega't}\hat{a}_{i}(\omega')\} \left(
  \matrix{
  \cos\theta_{2}\cr
  \sin\theta_{2}\cr
} \right),
\end{eqnarray}
while the negative frequency components are given by their
Hermitian conjugates. With these fields one can show that the
coincidence rate $N_{c}\propto\sin^{2}(\theta_{1}-\theta_{2})$
using the expressions developed in the Appendix.

Consider now that the sample, assumed to have
frequency-independent reflection coefficients, is placed in the
optical arrangement illustrated in Fig.~\ref{polariz-int}, and
that the polarizations of the downconverted photons are along the
eigenpolarizations of the sample. The effect of the sample, placed
in beam 1, may be represented by the following twin-photon Jones
matrix
\begin{equation}\label{jmatrix-sample}
 \textbf{T}_{s}=
  \left(\matrix{
    \textbf{R} & \textbf{0}\cr
    \textbf{0} & \textbf{I}\cr
  }\right),
  \end{equation}
  where
  \begin{equation}\label{sample-matrix}
  \textbf{R}=
\left (\matrix{
    \tilde{r}_{1} & 0\cr 0 & \tilde{r}_{2}\cr
    } \right)
  \end{equation}
(the justification for using this matrix to represent the action
of the sample is provided in the Appendix), $\textbf{I}$ is the
2$\times$2 identity matrix, and $\tilde{r}_{1}$ and
$\tilde{r}_{2}$ are the complex reflection coefficients of the
sample described earlier. The twin-photon Jones vector after
reflection from the sample and passage through the polarization
analyzers is given by
\begin{eqnarray}\label{jones-vect-sample-polarizer}
&\hat{\textbf{J}}_{3} &=
\textbf{T}_{p}\textbf{T}_{s}\hat{\textbf{J}}_{1}=\left( \matrix{
  j\{-\tilde{r}_{1}\cos\theta_{1}\hat{a}_{s}(\omega)+\tilde{r}_{2}\sin\theta_{1}\hat{a}_{i}(\omega')\}\left(
  \matrix{
  \cos\theta_{1}\cr
  -\sin\theta_{1}\cr
} \right)\cr
  \{\cos\theta_{2}\hat{a}_{s}(\omega)+\sin\theta_{2}\hat{a}_{i}(\omega')\}\left(
  \matrix{
  \cos\theta_{2}\cr
  \sin\theta_{2}\cr
} \right) } \right),
\end{eqnarray}
which results in
\begin{eqnarray}\label{field1-sample}
&\hat{\textbf{E}}_{1}^{+}(t)&=j\{-\tilde{r}_{1}\cos\theta_{1}\int
d\omega\, e^{-j\omega
t}\hat{a}_{s}(\omega)+\;\tilde{r}_{2}\sin\theta_{1}\int
d\omega'\,e^{-j\omega't}\hat{a}_{i}(\omega')\} \left(
  \matrix{
  \cos\theta_{1}\cr
  -\sin\theta_{1}\cr
} \right),
\end{eqnarray}
with $\hat{\textbf{E}}_{2}^{+}(t)$ identical to Eq.\
(\ref{field2-nosample}), since there is no sample in this beam.

Finally, using the expressions developed in the Appendix, it is
straightforward to show that
\begin{eqnarray}\label{coincidence-rate}
N_{c}=C[\tan^{2}\psi\cos^{2}\theta_{1}\sin^{2}\theta_{2}+\sin^{2}\theta_{1}\cos^{2}\theta_{2}-2\tan\psi\cos\Delta\cos\theta_{1}\cos\theta_{2}\sin\theta_{1}\sin\theta_{2}],
\end{eqnarray}
where the constant of proportionality $C$ depends on the
efficiencies of the detectors and the duration of accumulation of
coincidences. One can obtain $C$, $\psi$, and $\Delta$ with a
minimum of three measurements with different analyzer settings,
e.g., $\theta_{2}=0^{\circ}$, $\theta_{2}=90^{\circ}$, and
$\theta_{2}=45^{\circ}$, while $\theta_{1}$ remains fixed at any
angle except $0^{\circ}$ and $90^{\circ}$.

If the sample is replaced by a \textit{perfect} mirror, the
coincidence rate in Eq.\ (\ref{coincidence-rate}) becomes a
sinusoidal pattern of 100$\%$ visibility,
$C\sin^{2}(\theta_{1}-\theta_{2})$, as previously indicated. In
practice, by judicious control of the apertures placed in the
down-converted beams, visibilities close to 100$\%$ can be
obtained.

To understand the need for temporal compensation discussed
previously, we re-derive Eq.\ (\ref{coincidence-rate}), which
assumes full compensation, when a birefringent compensator is
placed in one of the arms of the configuration:
\begin{eqnarray}
N_{c}&=&C[\tan^{2}\psi\cos^{2}\theta_{1}\sin^{2}\theta_{2}+\sin^{2}\theta_{1}\cos^{2}\theta_{2}\nonumber\\&&-2\tan\psi\cos\Delta\cos\theta_{1}\cos\theta_{2}\sin\theta_{1}\sin\theta_{2}\Phi(\tau)\cos(\omega_{o}\tau)].
\end{eqnarray}
Here $\tau$ is the birefringent delay, $\omega_{o}$ is half the
pump frequency, and $\Phi(\tau)$ is the Fourier transform of the
SPDC normalized power spectrum. When $\tau=0$ we recover Eq.
(\ref{coincidence-rate}), whereas when $\tau$ is larger than the
inverse of the SPDC bandwidth, the third term that includes
$\Delta$ becomes zero and thus $\Delta$ cannot be determined.

The drawback of the arrangement illustrated in Fig. is the
requirement for a beam splitter, as in classical interferometric
ellipsometry. Any deviation from the assumed symmetric
reflectance/transmittance of this device will impair the
measurements and necessitate the use of a reference sample for
calibration.

\subsection{ENTANGLED TWIN-PHOTON ELLIPSOMETER}

As in classical interferometry, the configuration in the previous
section uses a beam splitter as a means of creating the multiple
paths that lead to interference. We now show that one can
construct an interferometer that makes use of quantum
entanglement, which then dispenses with the beam splitter. This
has the salutary effect of keeping 100$\%$ of the incoming photon
flux (rather than 50$\%$) while eliminating the requirement of
characterizing it. Moreover, no other optical elements are
introduced, so one need not be concerned with the characterization
of any components. This is a remarkable feature of
entanglement-based quantum interferometry.

The NLC is adjusted to produce SPDC in a type-II noncollinear
configuration, as illustrated in
Fig.~\ref{polariz-int-noncollinear}. Following the procedure
discussed in the previous section it is straightforward to show
that resulting coincidence rate is given by
\begin{eqnarray}\label{coincidence-rate-noncollinear}
N_{c}&=&C[\tan^{2}\psi\cos^{2}\theta_{1}\sin^{2}\theta_{2}+\sin^{2}\theta_{1}\cos^{2}\theta_{2}
+2\tan\psi\cos\Delta\cos\theta_{1}\cos\theta_{2}\sin\theta_{1}\sin\theta_{2}].
\end{eqnarray}
This expression is virtually identical to the one presented in
Eq.\ (\ref{coincidence-rate}) (except for the substitution of the
plus sign for the minus sign in the last term). An interesting
feature of this interferometer is that it is not sensitive to an
overall mismatch in the length of the two arms of the setup and
this increases the robustness of the arrangement.

An illuminating way of representing the action of the entangled
twin-photon quantum ellipsometer is readily achieved by redrawing
Fig.~\ref{polariz-int-noncollinear} in the unfolded configuration
shown in Fig.~\ref{unfolded}. Using the advanced-wave
interpretation, which was suggested by Klyshko in the context of
twin-photon imaging \cite{DK2}, the coincidence rate for photons
at D$_{1}$ and D$_{2}$ may be obtained by tracing light waves
originating from D$_{2}$ to the NLC and then onto D$_{1}$ upon
reflection from the sample. With this interpretation, the
configuration in Fig.~\ref{unfolded} becomes geometrically similar
to the classical ellipsometer. Although none of the optical
components usually associated with interferometers (beam splitters
and wave plates) are present in this scheme, interferometry is
still effected via the entanglement of the source.

An advantage of this setup over its idealized null ellipsometric
counterpart, discussed in section 2, is that the two arms of the
ellipsometer are separate and the light beams traverse them
independently in different directions. This allows various
instrumentation errors of the classical setup to be circumvented.
For example, placing optical elements before the sample causes
beam deviation errors \cite{JZ} when the faces of the optical
components are not exactly parallel. This leads to an error in the
angle of incidence and, consequently, errors in the estimated
parameters. In our case no optical components are placed between
the source (NLC) and the sample; any desired polarization
manipulation may be performed in the other arm of the entangled
twin-photon ellipsometer. Furthermore, one can change the angle of
incidence to the sample easily and repeatedly.

A significant drawback of classical ellipsometry is the difficulty
of fully controlling the polarization of the incoming light. A
linear polarizer is usually employed at the input of the
ellipsometer, but the finite extinction coefficient of this
polarizer causes errors in the estimated parameters \cite{RA}. In
the entangled twin-photon ellipsometer the polarization of the
incoming light is dictated by the phase-matching conditions of the
nonlinear interaction in the NLC. The polarizations defined by the
orientation of the optical axis of the NLC play the role of the
input polarization in classical ellipsometry. The NLC is aligned
for type-II SPDC so that only one polarization component of the
pump generates SPDC, whereas the orthogonal (undesired) component
of the pump does not (since it does not satisfy the phase-matching
conditions). The advantage is therefore that the downconversion
process assures the stability of polarization along a particular
direction.

\section{CONCLUSION}

Classical ellipsometric measurements are limited in their accuracy
by virtue of the need for an absolutely calibrated source and
detector. Mitigating this limitation requires the use of a
well-characterized reference sample in a null configuration.

Twin-photon ellipsometry, which makes use of simultaneously
emitted photon pairs, is superior because it removes the need for
a reference sample. Nevertheless, the unentangled twin-photon
ellipsometer requires that the optical components employed in the
interferometric arrangement be well characterized.

We have demonstrated that entangled twin-photon ellipsometry is
self-referencing and therefore eliminates the necessity of
constructing an interferometer altogether. The underlying physics
that leads to this remarkable result is the presence of
fourth-order (coincidence) quantum interference of the photon
pairs in conjunction with nonlocal polarization entanglement.

Our proposed entangled twin-photon ellipsometer is subject to the
same shot-noise-limited, as well as angularly resolved,
\textit{precision} that is obtained with traditional ellipsometers
(interferometric and null systems, respectively), but removes the
limitation in \textit{accuracy} that results from the necessity of
using a reference sample in traditional ellipsometers.

Since the SPDC source is inherently broadband, narrowband spectral
filters must be used to ensure that the ellipsometric data are
measured at a specific frequency. Spectroscopic data can be
obtained by employing a bank of such filters. Alternatively,
techniques from Fourier-transform spectroscopy may be used to
directly make use of the broadband nature of the source in
ellipsometric measurements.

\section*{APPENDIX}

\setcounter{equation}{0}
\renewcommand{\theequation}
      {\mbox{{A}{\arabic{equation}}}}

We investigate the effect that reflection from a sample has on the
quantized-field operators. We model the sample as a lossless beam
splitter, with complex reflection coefficient $\tilde{r}$, and
complex transmission coefficient $\tilde{t}$, that transforms the
input field operators $\hat{a}_{1}$ and $\hat{a}_{v}$ into output
field operators $\hat{b}_{1}$ and $\hat{b}_{v}$ according to
\begin{equation}\label{A1}
\hat{b}_{1}=j\tilde{r}\hat{a}_{1}+\tilde{t}\hat{a}_{v},\quad\hat{b}_{v}=\tilde{t}\hat{a}_{1}+j\tilde{r}\hat{a}_{v},
\end{equation}
where $|\tilde{t}|^{2}+|\tilde{r}|^{2}=1$ (so that the bosonic
commutation relations are preserved for $\hat{b}_{1}$ and
$\hat{b}_{v}$), $\hat{a}_{1}$ is the annihilation operator of a
single mode of the incident optical field, while $\hat{a}_{v}$ is
the annihilation operator of the vacuum entering the other port of
the beam splitter.

The coincidence rate at the detectors at times $t_{1}$ and $t_{2}$
is given by
\begin{equation}\label{A2}
G(t_{1},t_{2})=\langle\Psi|{\hat{E_{1}}}^{(-)}(t_{1}){\hat{E_{2}}}^{(-)}(t_{2}){\hat{E_{2}}}^{(+)}(t_{2}){\hat{E_{1}}}^{(+)}(t_{1})|\Psi\rangle,
\end{equation}
where $\hat{E_{1}}^{(+)}(t)=\int d\omega\, e^{-j\omega
t}\hat{b}_{1}(\omega)$, $\hat{E_{2}}^{(+)}(t)=\int d\omega\,
e^{-j\omega t}\hat{a}_{2}(\omega)$, and $|\Psi\rangle$ is the
twin-photon state at the output of the NLC:
\begin{eqnarray}\label{A3}
|\Psi\rangle=\int
d\omega\,\varphi(\omega,\omega_p-\omega)|1_\omega,1_{\omega_{p}-\omega},0_\omega\rangle.
\end{eqnarray}
The first element in the ket corresponds to the signal mode at
frequency  $\omega$, the second element corresponds to the idler
mode at frequency  $\omega_{p}-\omega$ (conservation of energy
ensures that the signal and idler frequencies add up to the pump
frequency $\omega_{p}$), while the third element corresponds to
the vacuum mode at frequency $\omega$ at the other input port of
the beam splitter that represents the sample. The function
$\varphi(\omega,\omega_p-\omega)$ is the probability amplitude of
the possible combinations of frequencies for pairs of signal and
idler modes emitted by the NLC.

Inserting the identity operator
$\sum_{n_{1},n_{2},n_{3}}|n_{1},n_{2},n_{3}\rangle\langle
n_{1},n_{2},n_{3}|$ (represented in the Fock basis of the Hilbert
space spanned by the signal, idler, and vacuum fields) into Eq.
(A2) gives
\begin{eqnarray}\label{A4}
G(t_{1},t_{2})&=&\sum_{n_{1},n_{2},n_{3}}\langle\Psi|{\hat{E_{1}}}^{(-)}(t_{1}){\hat{E_{2}}}^{(-)}(t_{2})|n_{1},n_{2},n_{3}\rangle
\langle
n_{1},n_{2},n_{3}|{\hat{E_{2}}}^{(+)}(t_{2}){\hat{E_{1}}}^{(+)}(t_{1})|\Psi\rangle\nonumber\\&
=&\sum_{n_{1},n_{2},n_{3}}|\langle
n_{1},n_{2},n_{3}|{\hat{E_{2}}}^{(+)}(t_{2}){\hat{E_{1}}}^{(+)}(t_{1})|\Psi\rangle|^{2}.
\end{eqnarray}
It is straightforward to show that using the state in Eq.\
(\ref{A3}) results in all terms in the summation in Eq.\
(\ref{A4}) vanishing except for the term where
$n_{1}=n_{2}=n_{3}=0$, so that
$G(t_{1},t_{2})=|\langle0,0,0|{\hat{E_{2}}}^{(+)}(t_{2}){\hat{E_{1}}}^{(+)}(t_{1})|\Psi\rangle|^{2}
$, and also results in the terms containing $\tilde{t}\hat{a}_{v}$
vanishing. Thus, in coincidence measurements, the effect of
reflection from the sample appears as a direct multiplication of
the relevant operators by the suitable complex reflection
coefficient, which justifies the use of the matrix in Eq.\
(\ref{sample-matrix}) when two orthogonal polarizations of the
fields (and thus two corresponding complex reflection
coefficients) are taken into consideration. Note that the
detectors actually record a time-averaged coincidence rate $N_{c}$
since the response time for optical detectors is usually much
longer than the inverse bandwidth of the the function
$\varphi(\omega,\omega_p-\omega)$ (see Ref. \cite{BS} for
details).

\section*{ACKNOWLEDGMENTS}

This work was supported by the National Science Foundation and by
the Center for Subsurface Sensing and Imaging Systems (CenSSIS),
an NSF engineering research center.

A. V. Sergienko's e-mail address is alexserg@bu.edu.


\newpage
\begin{figure}[hb]
\caption{The null ellipsometer. S is an optical source, P a linear
polarizer, $\frac{\lambda}{4}$ a quarter-wave plate (compensator),
A a linear polarization analyzer, and D an optical detector;
$\theta_i$ is the angle of incidence. The sample is characterized
by the ellipsometric parameters $\psi$ and $\Delta$ defined in the
text.}\label{null}
\end{figure}

\begin{figure}[hb]
\caption{Unentangled twin-photon ellipsometer. NLC stands for
nonlinear crystal; BS is a non-polarizing beam splitter; A$_1$ and
A$_2$ are linear polarization analyzers; D$_1$ and D$_2$ are
single-photon detectors; and $N_c$ is the coincidence
rate.}\label{polariz-int}
\end{figure}

\begin{figure}[hb]
\caption{Entangled twin-photon
ellipsometer.}\label{polariz-int-noncollinear}
\end{figure}

\begin{figure}[hb]
\caption{Unfolded version of the entangled twin-photon
ellipsometer displayed in Fig. 3.}\label{unfolded}
\end{figure}

\newpage

\begin{figure}[h]
 \epsfxsize=6.8in \epsfysize=5.2in
 \epsffile{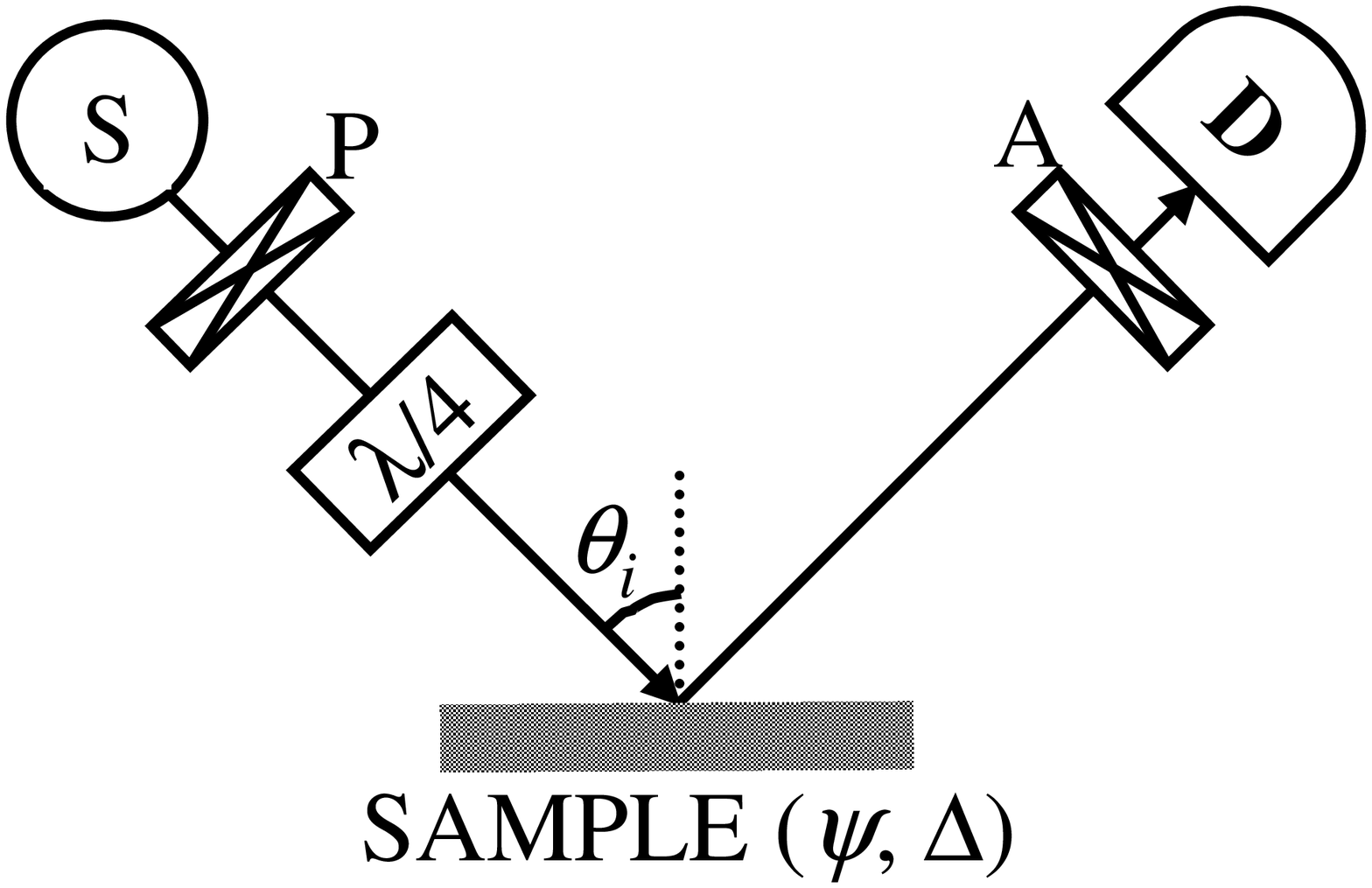}
 \end{figure}
\vskip6cm Figure 1, A. F. Abouraddy

\begin{figure}[h]
 \epsfxsize=6.8in \epsfysize=5.2in
 \epsffile{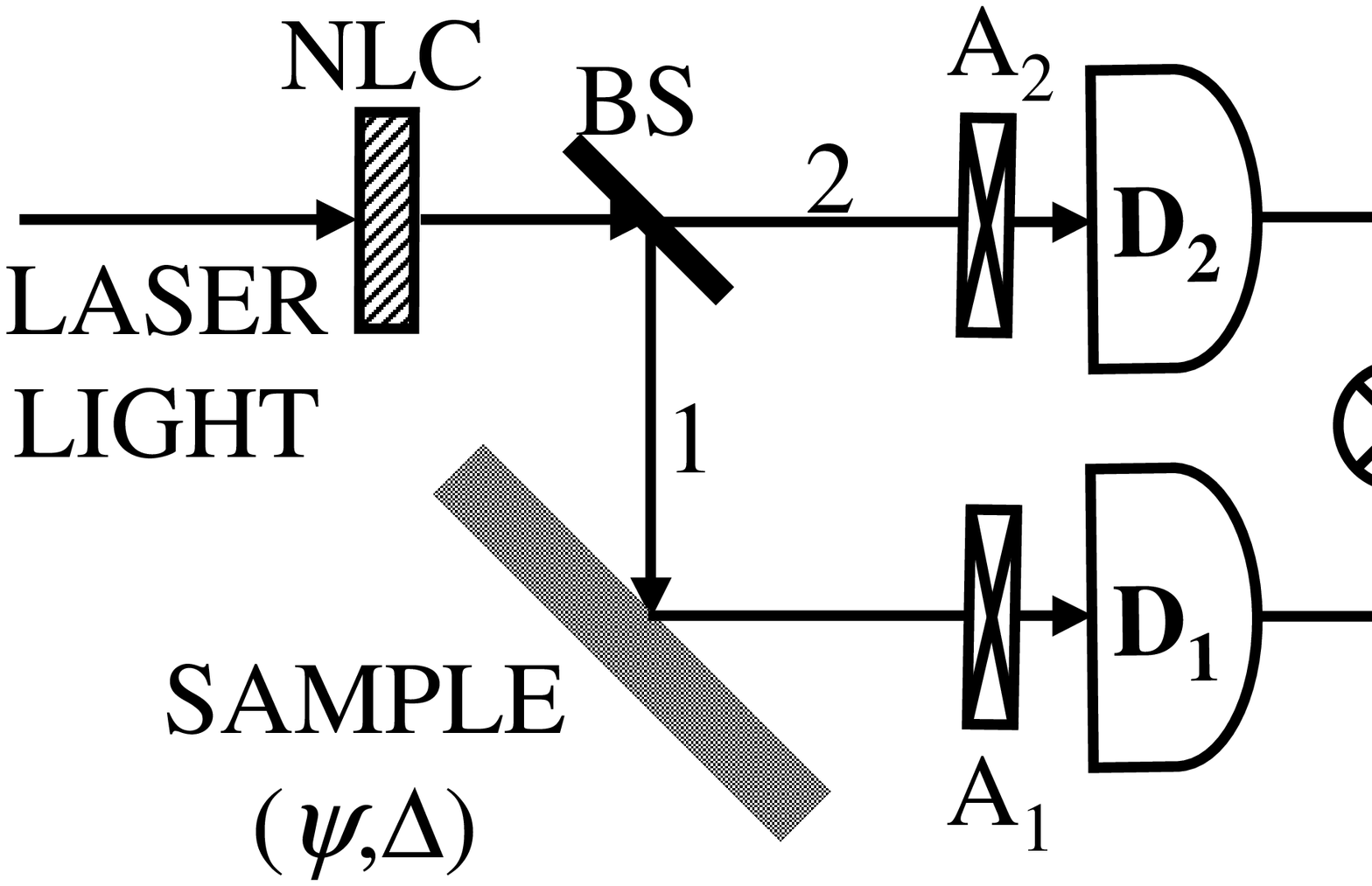}
 \end{figure}
\vskip6cm Figure 2, A. F. Abouraddy

\begin{figure}[h]
 \epsfxsize=6.8in \epsfysize=5.2in
 \epsffile{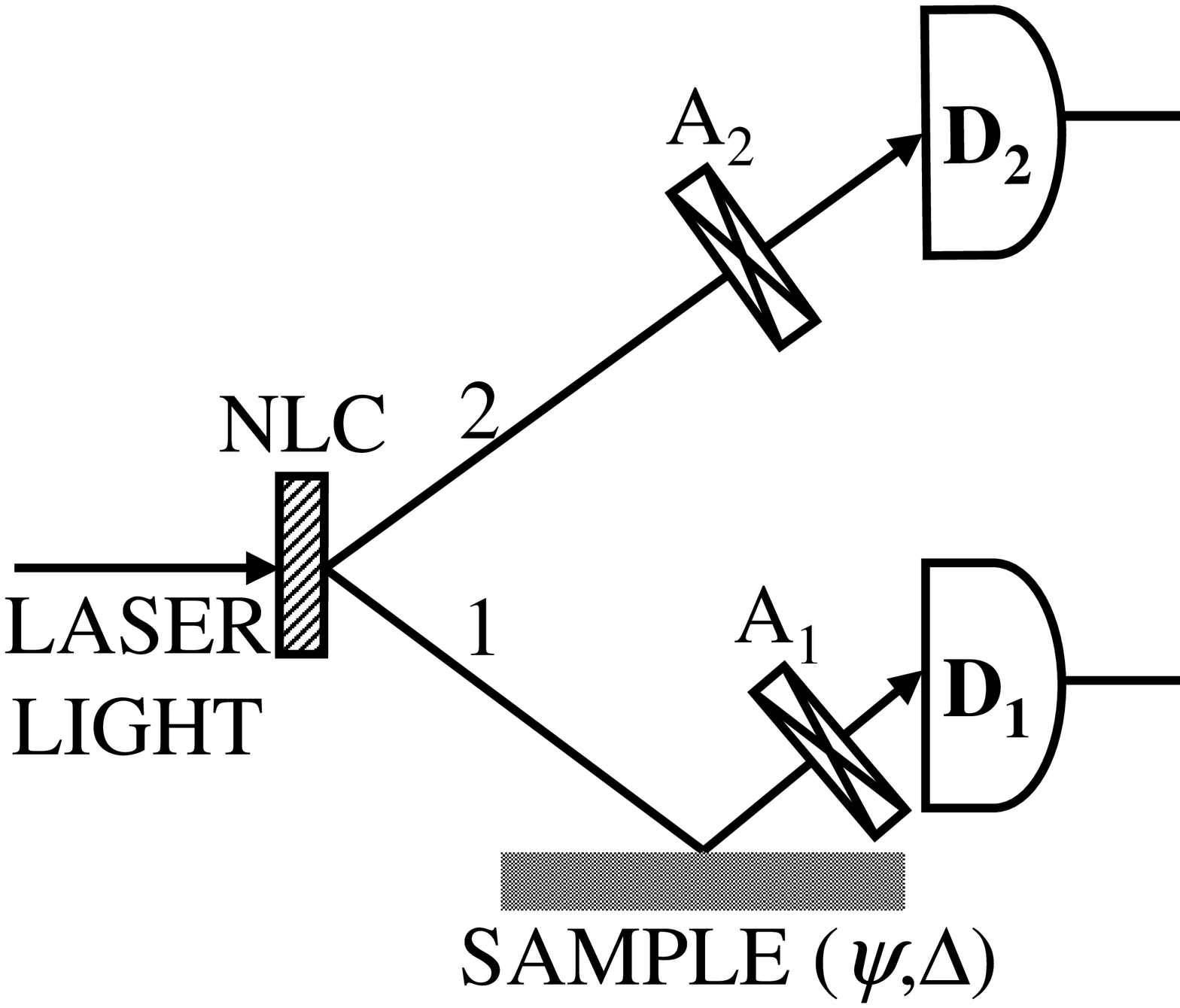}
 \end{figure}
\vskip6cm Figure 3, A. F. Abouraddy

\begin{figure}[h]
 \epsfxsize=6.8in \epsfysize=5.2in
 \epsffile{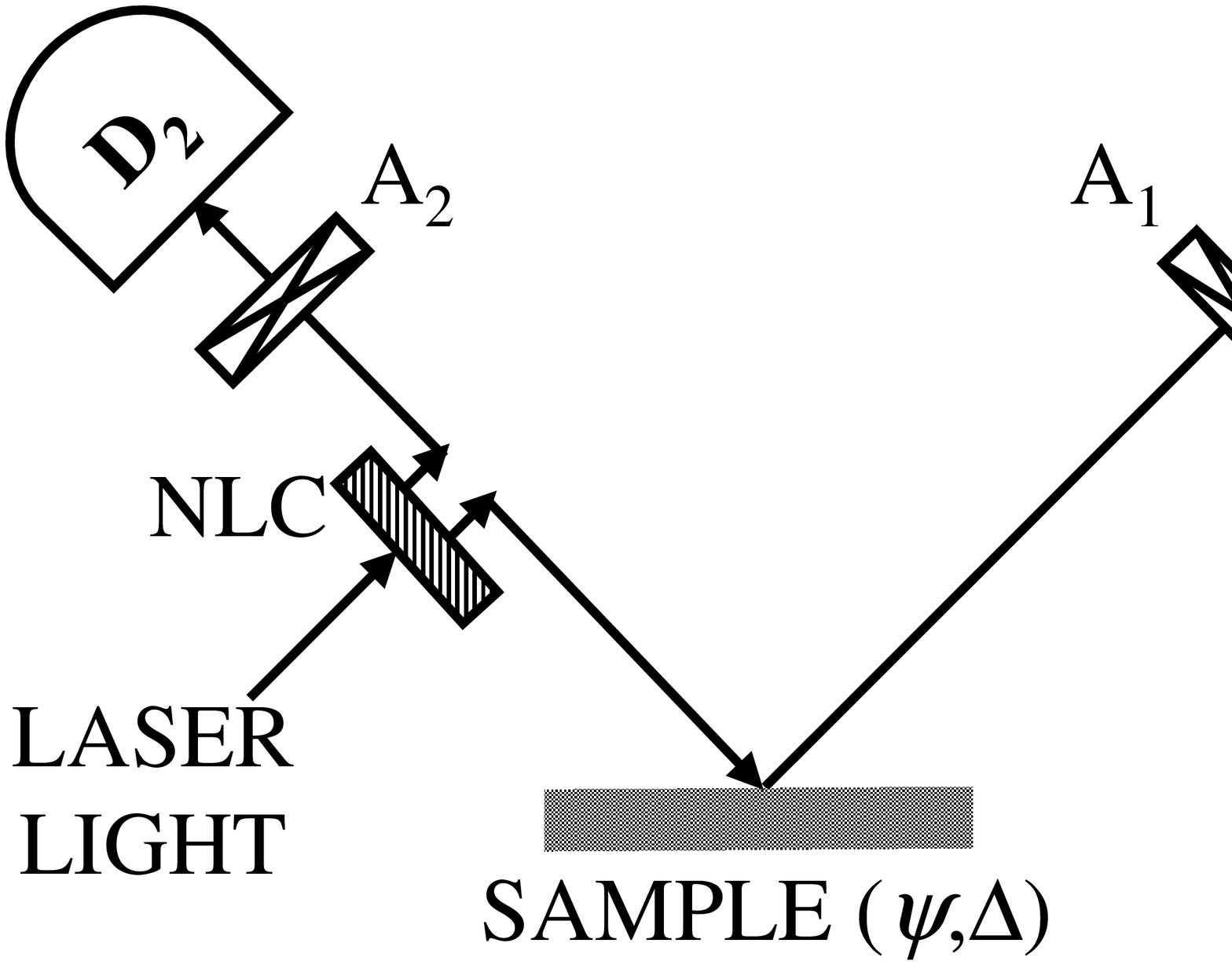}
 \end{figure}
\vskip6cm Figure 4, A. F. Abouraddy

\end{document}